\documentclass[twocolumn,prl,showpacs]{revtex4}
\usepackage{graphicx}
\usepackage{dcolumn}
\usepackage{amsmath}
\usepackage{amssymb}
\usepackage{mathrsfs}
\usepackage{bm}

\newcommand{\kk}{\mathbf{k}}       

\begin{document}
\title{Quantum theory of photonic crystal polaritons}

\author{D. Gerace}
\author{M. Agio}
\author{L. C. Andreani}

\affiliation{Istituto Nazionale per la Fisica della Materia and
Dipartimento di Fisica ``Alessandro Volta'',
Universit\`a di Pavia, via Bassi 6, 27100 Pavia, Italy}

\begin{abstract}

We formulate a full quantum mechanical theory of the interaction 
between electromagnetic modes in photonic crystal slabs and quantum 
well excitons embedded in the photonic structure. We apply
the formalism to a high index dielectric layer with a periodic 
patterning suspended in air. 
The strong coupling between electromagnetic modes
lying above the cladding light line 
and exciton center of mass 
eigenfunctions manifests itself with the typical anticrossing 
behavior. The resulting band
dispersion corresponds to the quasi-particles coming from
the mixing of electromagnetic and material excitations, which we call
\textit{photonic crystal polaritons}. We compare the results
obtained by using the quantum theory to variable angle reflectance 
spectra coming from a scattering matrix approach, and 
we find very good quantitative agreement.  

\end{abstract}

\pacs{42.50.Ct, 42.70.Qs, 71.36.+c, 73.20.Mf, 78.20.Bh}

\maketitle                   

\textbf{Introduction}. 
Since the pioneering works of more than fifteen years ago by
E.~Yablonovitch and S.~John~\cite{yablojohn87},
a great deal of research in physics has been dedicated
to the study of the optical properties of photonic crystals~\cite{review}.
In the last few years much attention has been devoted to photonic 
crystal (PC) structures embedded in planar dielectric waveguides, which 
are also called {\it photonic crystal slabs}. These are systems with 
a periodic dielectric response in the plane of the waveguide, while a 
dielectric mismatch is used to confine the electromagnetic field in 
the vertical direction.\\ 
One of the main issues of PC slabs is the 
existence of the light line problem: only photonic modes lying below
the cladding light line are truly guided and with zero intrinsic linewidth,
while the modes lying above the light line are radiative. 
A method to calculate the
photonic band dispersion and the intrinsic linewidth (complex energies) both
below and above the light line has been recently
proposed~\cite{andreani_IEEE,andreani_pss}. Until now, however, 
the physics of PCs has been considered mostly for what concerns the 
``optical'' point of view, that is studying the propagation of light 
in periodic dielectric media.\\
In this work we analyze the radiation-matter interaction in PC
slabs, by considering the effects of the interplay between the 
electromagnetic field and semiconductor quantum well (QW) excitons. 
In particular, we show that polaritonic
effects are present in PC slabs when the exciton-photon coupling is 
larger than the intrinsic radiative linewidth of a photonic mode above
the cladding light line. The band dispersion is greatly modified in
the vicinity of the excitonic resonance, leading to the formation of
mixed states which are analogous to exciton-polaritons in bulk 
semiconductors~\cite{hopfield} and 
microcavities~[6--8] and 
which we call {\it photonic crystal polaritons} (PCPs).
Experimental results showing the formation of PCPs in organic-based 
systems were previously reported in Ref.~\onlinecite{fujita98}.    

\begin{figure*}[t]
\begin{center}
\includegraphics[width=\textwidth]{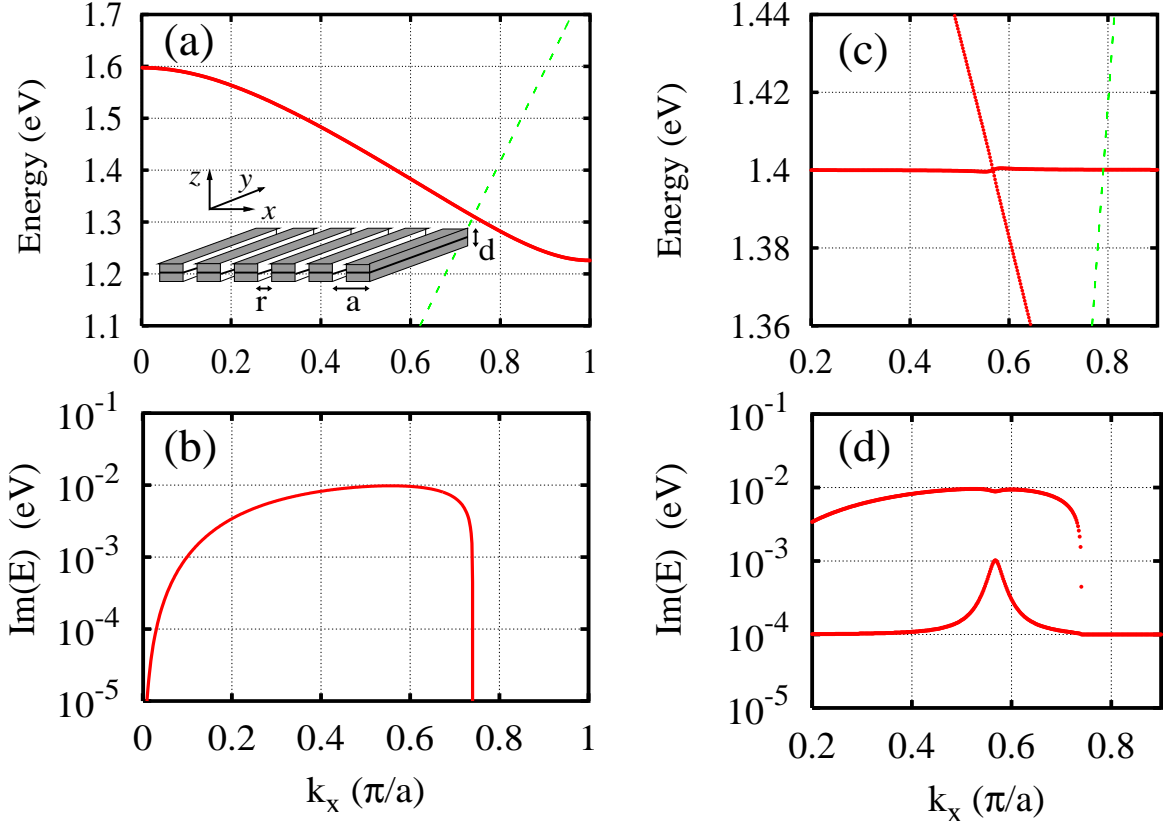}
\caption{(a) Real and (b) imaginary parts of the complex
           photonic energies of the even TE mode as a function
           of the wave vector in the first Brillouin zone,
           for the system shown in
           the inset, without excitonic resonance (no QW);
         (c) real and (d) imaginary parts for the same structure with
           a QW at the center of the dielectric core, with an excitonic
           resonance at $\hbar\Omega_0=1.4$~eV. The dashed lines in (a) and
           (c) represent the light line in air. }
\end{center}
\end{figure*}

\begin{figure*}[t]
\begin{center}
\includegraphics[width=\textwidth]{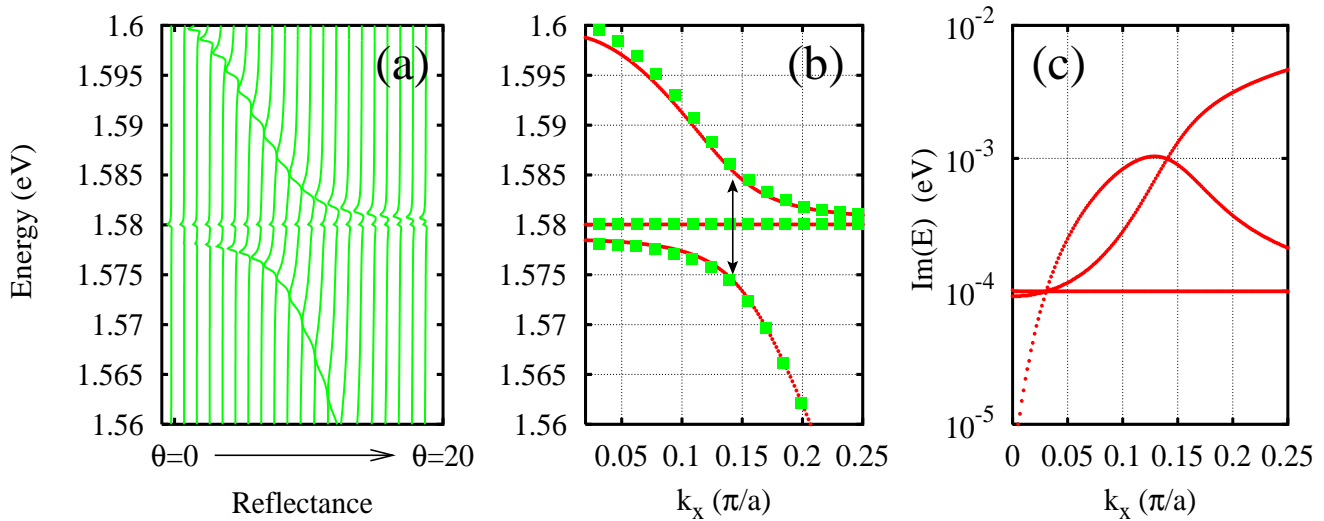}
\caption{ (a) Calculated surface reflectance of a TE polarized
              plane wave incident on the structure of the
              inset of Fig.~1a. The excitonic resonance is set at
              $\hbar\Omega_0=1.58$~eV; the angle of incidence is
              varied from $\theta=0^{\circ}$ to $\theta=20^{\circ}$,
              with a step of $1^{\circ}$.
           (b) Real and (c) imaginary parts of the complex
              energies of the photonic crystal polaritons, calculated with the
              quantum theory. The squared points in (b) are extracted from
              the calculated reflectance curves shown in (a), and the
              polariton splitting is indicated by an arrow.   }
\end{center}
\end{figure*}

\textbf{Theory}. 
Here we briefly describe the quantum theory of PCPs, 
starting with the second quantized total hamiltonian of the system, 
which is given by

\begin{small}
\begin{eqnarray}
\label{hamiltonian}
\hat{H} &=& \sum_{\kk,n}\hbar\omega_{\kk n} \hat{a}_{\kk n}^{\dagger}
          \hat{a}_{\kk n}
        + \sum_{\kk,\nu}\hbar\Omega_{\kk \nu}\hat{b}_{\kk \nu}^{\dagger}
          \hat{b}_{\kk \nu}\nonumber\\
        &+& i\sum_{\kk,n,\nu}C_{\kk n \nu}
                 (\hat{a}_{\kk n}+\hat{a}_{-\kk n}^{\dagger})
                 (\hat{b}_{\kk \nu}^{\dagger}-\hat{b}_{-\kk \nu})\nonumber\\ 
        &+& \sum_{\kk,\nu,n_1,n_2}\frac{C_{\kk n_1 \nu}^{*}
       C_{\kk n_2 \nu}}{\hbar\Omega_{\kk \nu}}
       (\hat{a}_{-\kk n_1}+\hat{a}_{\kk n_1}^{\dagger})
       (\hat{a}_{\kk n_2}+\hat{a}_{-\kk n_2}^{\dagger})\quad .
\end{eqnarray}
\end{small}  

In Eq.~(\ref{hamiltonian}), the first term indicates the photonic band 
dispersion (real part of the complex eigenenergies), 
$\hat{a}_{\kk n}$ ($\hat{a}_{\kk n}^{\dagger}$) being the destruction 
(creation) operators of a photon with wave vector $\kk$ and band number $n$. 
These energies are obtained by expanding the magnetic field in terms of the
guided modes of the effective waveguide, that is the one with an average 
dielectric constant, and then by solving the Maxwell equation as a linear
eigenvalue problem~\cite{andreani_IEEE}. The imaginary part of the photonic
modes is calculated by using a perturbative approach~\cite{andreani_pss}. 
If we restrict our considerations to the system 
schematically shown in the inset of Fig.~1a, the direction of periodicity is
$x$ and thus the wave vector is given by the component $k_x$. The modes in 
a PC slab suspended in air can be classified as even (odd) with respect 
to specular reflection through the plane $xy$, and even (odd) with 
respect to the plane of incidence (that is $xz$ in the case considered here). 
Throughout this paper we consider only modes which are spatially even 
with respect to $xy$ plane, and odd with respect to the vertical mirror 
plane (even TE modes). These modes interact with transverse QW excitons.
The excitonic problem, whose solution gives the energies
$\hbar\Omega_{\kk \nu}$, is treated by solving the Schr\"odinger 
equation for the exciton center of mass envelope function in a 
periodic piecewise constant potential having the same patterning as
the photonic crystal structure. The in-plane 
wave vector and an integer $\nu$ are good quantum numbers for the 
exciton wavefunctions, due to the periodic potential and to the spatial
dispersion of the center of mass; the corresponding destruction (creation)
operators are $\hat{b}_{\kk\nu}$ ($\hat{b}_{\kk\nu}^{\dagger}$).  
The exciton-photon coupling matrix elements, $C_{\kk n \nu}$, are calculated
in terms of the microscopic physical quantities as
\begin{equation}
\label{coupling}
C_{\kk n\nu} =  \left(\frac{2\pi e^2\hbar\Omega_{\kk\nu}^2}
                {\omega_{\kk n}}\right)^{1/2}\langle
                \Psi^{(exc)}_{\kk\nu}|\sum_{j}\mathbf{E}_{\kk n}
                (\mathbf{r}_{j})\cdot \mathbf{r}_{j}|0\rangle\,\,\, .
\end{equation}
 
In Eq.~(\ref{coupling}), $\Psi^{(exc)}_{\kk\nu}$ is the all-electron
exciton wavefunction, $\mathbf{E}_{\kk n}$ is the electric field
profile for the photonic mode at frequency $\omega_{\kk n}$
in the PC slab, and the sum is over all the QW electrons. 
In particular, $C_{\kk n \nu}$ is found to depend on the overlap
between the exciton envelope function and the transverse
electric field (for what concerns TE modes). It is proportional to
$(f/S)^{1/2}$, where $f/S$ is the oscillator strength
per unit area which depends on the QW thickness~\cite{andreani_sif}. 
The hamiltonian is diagonalized by using a generalized Hopfield 
transformation to
expand new destruction (creation) operators $\hat{P}_{\kk}$ 
($\hat{P}_{\kk}^{\dagger}$) as a linear combination of $\hat{a}_{\kk n}$ 
($\hat{a}_{\kk n}^{\dagger}$) and $\hat{b}_{\kk \nu}$ 
($\hat{b}_{\kk \nu}^{\dagger}$), with the condition
$[\hat{P}_{\kk},\hat{H}]=E_{\kk}\hat{P}_{\kk}$~\cite{giovanna99}. 
The new eigenenergies $E_{\kk}$
correspond to mixed excitations of radiation and matter, i.~e. the PCPs.

All the results presented in this paper refer to a model structure,
namely a high index dielectric
core suspended in air with a one dimensional patterning,
and a QW placed at the center of it (see inset of Fig.~1a).
The dielectric constant of the semiconductor-based core layer is set
to the value $\varepsilon=12$, the dielectric constant of the air being
simply $\varepsilon_{air}=1$. The slab
thickness and the air fraction are set to the values
$d/a=0.2$ and $r/a=0.3$ respectively, the lattice constant 
being $a=350$~nm.
A QW of thickness $L_{QW}=8$~nm is considered, and the oscillator strength
per unit area is set to the typical value $f/S=8.4\times 10^{12}$~cm$^{-2}$.
The intrinsic radiative exciton linewidth is assumed to be 
$\Gamma=0.1$~meV.

\textbf{Results and Discussion}.
In Fig.~1a we show the photonic band dispersion  
in the energy range 1.1-1.7~eV. In Fig.~1b we display the 
corresponding imaginary part as a function of the wave vector, which
shows a maximum at about 1.4~eV. The imaginary part goes to zero at
$k_x=0$, that is at normal incidence, and at $k_x=0.74$; this last 
behavior is due to the crossing of the light line, as shown in Fig.~1a,
corresponding to the photonic mode becoming truly guided and stationary.  
If the photonic imaginary part is larger than the exciton-photon 
coupling matrix element (which is of the order of a few meV), 
a QW placed at the center of the PC slab does not
produce any important change in the photonic band dispersion. Indeed, this 
result is shown in Figs.~1c and 1d, in which we display the real and 
imaginary parts of the complex eigenenergies coming from the 
diagonalization of Eq.~(\ref{hamiltonian}) with an excitonic
resonance at $\hbar\Omega_0=1.4$~eV. 
In this \textit{weak coupling} regime
the photon and the exciton are almost uncoupled, as it can be seen f
rom the crossing of the two dispersion relations in Fig.~1c. 
The imaginary part of the exciton increases by an order of magnitude
correspondingly to the crossing point, but this has a negligible effect
on the photonic radiative linewidth, which is still an order of magnitude 
larger than the excitonic one (see Fig.~1d).\\
In order to observe the \textit{strong coupling} regime, 
the energy of the excitonic resonance has to lie where the imaginary
part of the photonic mode is smaller than the coupling matrix element.
From Fig.~1a and 1b we see that for $\hbar\Omega_0=1.58$~eV the imaginary part
of the corresponding photonic band is about Im$(\hbar\omega)=10^{-3}$~eV.
In Fig.~2a we show the results of a variable angle reflectance calculation, 
which is done by a scattering matrix approach~\cite{whittaker99}. 
The dielectric function of the QW layer is frequency dependent, 
with a resonance at $\hbar\Omega_0=1.58$~eV.  
It is known that the sharp
features appearing in the reflectance spectrum correspond to the 
excitation of photonic modes above the light 
line~\cite{fujita98,astratov99}, thus giving
a point $(\kk,\omega)$ of the corresponding photonic band 
dispersion; the wave vector component parallel to the surface is
given by $k=(\omega/c)\sin\theta$. The $(\kk,\omega)$ 
points extracted from the results of Fig.~2a are compared to the 
dispersion of PCPs in Fig.~2b, in which the squares are the
scattering matrix results. The quantum theory
is in excellent agreement with the classical approach. 
In particular, both theories confirm the 
anticrossing behavior of exciton and photon modes, which is a clear
effect of the strong coupling regime. With the parameters used in this work, 
the polariton splitting is as high as 10~meV at the anticrossing point 
$k_x=0.14$, as shown in Fig.~2b. This splitting is found to be slightly
larger than in semiconductor-based 
microcavities~\cite{skolnik98,andreani_sif}.
In Fig.~2c, finally, the imaginary part of the PCPs complex
eigenenergies is shown. At $k_x=0.14$ the imaginary parts of
the upper and lower polariton branches become equal to the same value
Im$(E)=10^{-3}$~eV, thereby indicating that
mixed states of radiation and matter form in the PC slab. 
The dispersionless curve at Im$(E)=10^{-4}$~eV
in Fig.~2c corresponds to the uncoupled excitonic modes. 

\textbf{Conclusions}.
We have shown that the quantum theory of the interaction
between electromagnetic modes in photonic crystal slabs and QW excitons
can describe both the weak and strong coupling regimes. Moreover,
we have found a very good agreement between the quantum theory and 
a classical approach based on a scattering matrix method. We thus
conclude that new mixed states of radiation and matter excitations can
be experimentally measured in semiconductor-based photonic crystal slabs
by using variable angle reflectance or transmittance techniques, 
provided that the experimental conditions required for being in the
strong coupling regime could be satisfied.

\textbf{Acknowledgements}.
This paper is dedicated to the memory of Giovanna Panzarini, whose 
research on exciton polaritons in 3D microcavities was of great
inspiration for the present developments. This work was supported 
by MIUR through Cofin 2002, and by INFM through PRA PHOTONIC.

\end{document}